\documentclass[final,5p,times,twocolumn]{elsarticle}

\biboptions{sort&compress}

\usepackage{graphicx}
\usepackage{amssymb}
\usepackage{amsmath}
\usepackage{bm}

\graphicspath{{.}{./EPS/}}

\newcommand* {\vek}[1]{{\ensuremath{\bm{\mathrm{#1}}}}}
\newcommand* {\ee}{\ensuremath{\mathrm{e}}}
\newcommand{\ket}[1]{\left | \, #1 \right \rangle}
\newcommand{\bra}[1]{\left \langle #1 \, \right |}

\journal{Physica E}

\begin{document}

\begin{frontmatter}



\title{AC transport properties of single and bilayer graphene}


\author[ifs]{J.~Z. Bern\'ad}
\author[ifs,ctcp]{U. Z\"ulicke\corref{contact}}
\author[aug]{K. Ziegler}

\address[ifs]{Institute of Fundamental Sciences and MacDarmid Institute for Advanced
Materials and Nanotechnology, Massey University (Manawatu Campus), Private Bag
11~222, Palmerston North 4442, New Zealand}
\address[ctcp]{Centre for Theoretical Chemistry and Physics, Massey University
(Albany Campus), Private Bag 102904, North Shore MSC, Auckland 0745, New
Zealand}
\address[aug]{Institut f\"ur Physik, Universit\"at Augsburg, D-86135 Augsburg, Germany}

\cortext[contact]{Corresponding author. Tel.: +64 6 356 9099 x7259.
E-mail address: u.zuelicke@massey.ac.nz (U. Z\"ulicke).}

\begin{abstract}
We have performed a theoretical study of electronic transport in single and
bilayer graphene based on the standard linear-response (Kubo) formalism and
continuum-model descriptions of the graphene band structure. We are focusing
especially on the interband contribution to the optical conductivity $\sigma(\omega)$.
Analytical results are obtained for a variety of situations, which allow clear 
identification of features in $\sigma(\omega)$ that are associated with relevant 
electronic energy scales. Our work extends previous numerical studies and
elucidates ways to infer electronic properties of graphene samples from
optical-conductivity measurements.
\end{abstract}

\begin{keyword}
graphene \sep optical conductivity \sep interband transitions
\PACS 81.05.Uw \sep 72.10.Bg \sep 73.23.Ad \sep 73.43.Cd


\end{keyword}

\end{frontmatter}


\section{Introduction}

Graphene is a single sheet of carbon atoms tightly packed into a two-dimensional
(2D) honeycomb lattice. This material has recently become available for experimental
study~\cite{novo:sci:04,novo:pnas:05}, and its exotic electronic properties are
attracting a lot of interest~\cite{castro:rmp:09}. In particular, the conical shape of
conduction and valence bands, together with the absence of a gap, near the
$\vek{K}$ and $\vek{K^\prime}$ points in the Brillouin zone renders graphene an
intriguing type of quasi-relativistic condensed-matter system~\cite{kats:ssc:07}. 
Recent experiments have verified that the band dispersion of charge carriers in
graphene is indeed linear as expected for massless Dirac fermions~\cite{novo:nat:05,
kim:nat:05}. A multitude of interesting physical effects arising in single-layer and
bilayer graphene samples have been discussed~\cite{castro:rmp:09}, both theoretically
and experimentally.

In this work, we focus on the ac electric transport properties of graphene, which have
been the subject of numerous theoretical (mostly numerical)
studies~\cite{kat:epjb:06,gus:prl:06,gus:prb:06,gus:prl:07,gus:prb:07,ziegler:prl:06,ziegler:prb:07,peres:prb:06,nilsson:prl:06,nilsson:prb:08,abergel:prb:07,cserti:prb:07,cserti:prl:07,falk:epjb:07,snyman:prb:07,ryu:prb:07,nicol:prb:08,staub:prb:08,peres:ijmpb:08,peres:epl:08,falk:08,min:prl:09}
and several recent experiments~\cite{wang:sci:08,nair:sci:08,li:natphys:08,mak:prl:08,zhang:prb:08,li:prl:09}. Measurements of quantities related to the optical conductivity
are expected to give deeper insight into the electronic properties of graphene samples,
making it possible to infer details of their morphology~\cite{zhang:prb:08,min:prl:09}
and suitability for applications. It is thus important to obtain a clear understanding of the
features exhibited in the frequency-dependent conductivity, in particular, their relation to
microscopic parameters and behavior at finite temperature $T$. Furthermore, it is
advantageous to have mathematical expressions available that can be straightforwardly
used for comparison with data. This is the motivation for our study. We have developed
a formalism that lends itself for generalization to many situations, in particular, the
treatment of inelastic scattering. Extrapolation to zero frequency will enable us to
discuss the dc conductivity, shedding new light on the phenomenon of minimal
conductivity in graphene. The full range of our results will be presented elsewhere;
here we focus on discussing the method and present selected results.

\section{Calculation of the conductivity: Basic formalism}
\label{sec:basics}

Our starting point is the familiar~\cite{madelung} Kubo formula
\begin{equation}\label{kuboStart}
\sigma_{\mu\nu} (\omega) = \int_{-\infty}^0 \!\! d t \,\,\, \ee^{i ( \omega - i 0^+ ) t} \,\,
K_{\mu\nu} \quad ,
\end{equation}
with the kernel
\begin{equation}\label{kernel1}
K_{\mu\nu} = \frac{i e}{\hbar} \, \text{Tr} \left\{ \ee^{-i \frac{H t}{\hbar}} j_\mu \ee^{i
\frac{H t}{\hbar}} \left[ r_\nu \, , \, \varrho \right] \right\} \quad .
\end{equation}
Here $j_\mu = - e\dot{r}_\mu \equiv \frac{-e}{\hbar}\, i [H\, , \, r_\mu]$ is the current
operator, and $\varrho$ the density matrix. An alternative expression for the
kernel~\cite{madelung},
\begin{equation}\label{kernel2}
K_{\mu\nu} =  \int_0^{\frac{1}{k_{\text{B}} T}} \!\!d\lambda \,\,\,  \text{Tr} \left\{ \ee^{-i
\frac{H t} {\hbar}} j_\mu \ee^{i \frac{H t}{\hbar}} \, \ee^{-\lambda H} j_\nu
\ee^{-\lambda H} \, \varrho \right\} \quad ,
\end{equation}
will become particularly useful to enable discussion of the effect of inelastic scattering.
When quasiparticle interactions are neglected, it is possible~\cite{ziegler:prl:06} to
express the conductivity in terms of matrix elements between eigenstates $\ket{n}$ of
the single-particle Hamiltonian having energy $\epsilon_n$:
\begin{eqnarray}\label{KlausConduct}
\sigma_{\mu\nu}(\omega) &=& \frac{e^2}{i\hbar}\sum_{n,n^\prime} \frac{\bra{n} [H\, , \, 
r_\mu] \ket{n^\prime} \bra{n^\prime}[H\, , \, r_\nu] \ket{n}}{(\varepsilon_{n^\prime} -
\varepsilon_n)(\varepsilon_{n^\prime} -\varepsilon_n + \hbar\omega - i 0^+)} \nonumber
\\ && \hspace{3cm} \times \left[ f(\varepsilon_n) - f(\varepsilon_{n^\prime}) \right] .
\end{eqnarray}
Here $f(\varepsilon)=1/(1+\exp\{[\varepsilon - \mu]/[k_{\text{B}}T]\}$ denotes the Fermi
function, which depends on the chemical potential $\mu$. In the following, we use the
expression~(\ref{KlausConduct}) to derive conductivity formulae applicable to
graphene.

\subsection{Clean limit: Plane-wave representation}

Using continuum-model descriptions of the band structure near the $\vek{K}$-points,
single-particle eigenstates of clean graphene systems can be written as a direct
product of a plane wave in real space and a $2 N$-spinor (the latter subsuming the
two sub-lattice and $N$ layer-index degrees of freedom): $\ket{n} = \ket{\vek{k}}
\otimes\ket{\sigma}_{\vek{k}}$. In single-layer graphene, $\sigma$ distinguishes the
two (electron and hole) bands. Note that the spinor wave function depends on wave
vector $\vek{k}$. The current operator $j_\mu$ or, equivalently, the commutator
$[H\, , \, r_\mu]$, is diagonal in the real-space part $\ket{\vek{k}}$ while having
off-diagonal matrix elements in pseudospin space. It is then straightforward to
specialize Eq.~(\ref{KlausConduct}) to this case, finding
\begin{eqnarray}
\sigma_{\mu\nu} &=& \frac{e^2}{i\hbar} \sum_{\sigma,\sigma^\prime}^{2 N} \int \!\!
\frac{d^2k}{(2\pi)^2} \frac{\bra{\sigma}w_\mu(\vek{k})\ket{\sigma^\prime}_{\vek{k}}
\bra{\sigma^\prime}w_\nu(\vek{k})\ket{\sigma}_{\vek{k}}}{\varepsilon_{\vek{k}
\sigma^\prime} -\varepsilon_{\vek{k}\sigma} + \hbar\omega - i 0^+} \nonumber \\
&& \hspace{2cm} \times \frac{f(\varepsilon_{\vek{k}\sigma}) - f(\varepsilon_{\vek{k}
\sigma^\prime})}{\varepsilon_{\vek{k}\sigma^\prime} - \varepsilon_{\vek{k}\sigma}}
\quad .
\end{eqnarray}
Here $w_\mu(\vek{k})$ is found from $[H\, , \, r_\mu] \ket{\vek{k}} = w_\mu(\vek{k})
\ket{\vek{k}}$. Two contributions to the conductivity can be distinguished, arising
from terms with $\sigma=\sigma^\prime$ and $\sigma\ne\sigma^\prime$, respectively.
It is customary to call these the {\em intra\/}-band and {\em inter\/}-band contributions.
Defining $w_\mu^{\sigma\sigma^\prime}(\vek{k}) = \bra{\sigma}w_\mu(\vek{k})
\ket{\sigma^\prime}_{\vek{k}}$, we find
\begin{eqnarray}\label{intraTerm}
\frac{\sigma_{\mu\nu}^{\text{(intra)}}}{\sigma_0} &=& \frac{\delta(\hbar\omega)}{2}
\sum_{\sigma}^{2N} \int \!\! d^2 k\,\,\, w_\mu^{(\sigma\sigma)}(\vek{k}) \, w_\nu^{(
\sigma\sigma)}(\vek{k}) \, f^\prime(\varepsilon_{\vek{k}\sigma}) \, ,  \nonumber \\ \\
\label{interTerm}
\frac{\sigma_{\mu\nu}^{\text{(inter)}}}{\sigma_0} &=& \frac{\sinh\left(\frac{\hbar\omega}
{2 k_{\text{B}}T} \right)}{2 \hbar\omega} \sum_{\sigma\ne\sigma^\prime} \int \!\! d^2 k
\,\, \delta\left(\hbar \omega - \left[ \varepsilon_{\vek{k}\sigma} - \varepsilon_{\vek{k}
\sigma^\prime}\right] \right) \nonumber \\
&& \times\frac{- w_\mu^{(\sigma \sigma^\prime)}(\vek{k}) \,\, w_\nu^{(\sigma^\prime \!
\sigma)}(\vek{k})}{\cosh\left(\frac{\hbar\omega}{2k_{\text{B}}T} \right) + \cosh\left( \frac
{ \varepsilon_{\vek{k}\sigma} + \varepsilon_{\vek{k}\sigma^\prime} - 2 \mu}{2
k_{\text{B}}T}\right)} \, .
\end{eqnarray}
For brevity, we use the scale factor $\sigma_0=g e^2/(2\pi\hbar)$, where $g=4$ has
been introduced to account for the quasiparticle degeneracy (real spin and valley) in
graphene. The intra-band term (\ref{intraTerm}) is the usual dc Drude conductivity,
which depends on states in the vicinity of the Fermi surface where the derivative
$f^\prime$ of the Fermi function is peaked. It vanishes at the neutrality point ($\mu=0$)
in the zero-temperature limit. The inter-band contribution (\ref{interTerm}) is the
interesting part for finite $\omega$. It is calculated straightforwardly using
continuum-model descriptions of single-layer and bilayer graphene. The expression
(\ref{interTerm}) given here is well-suited for obtaining analytical results for the
dependence on temperature but cannot be used to go beyond the clean limit. To
discuss the effect of disorder, a more general formula is needed that will be given in
the next subsection.

\subsection{General conductivity formula in terms of Greens functions}

Mathematical manipulation of Eq.~(\ref{KlausConduct}) yields the conductivity
expressed in terms of single-particle Greens
functions~\cite{ziegler:prl:06,ziegler:prb:07,ryu:prb:07}. It has the general form
\begin{equation}\label{eq:newConduct}
\frac{\sigma_{\mu\nu}(\omega)}{\sigma_0} = \int_{-\infty}^{\infty} \!\!\! d \varepsilon \,\, 
{\mathcal T}_{\mu\nu}(\varepsilon, \omega) \left[ f\left(\varepsilon+\frac{\hbar\omega}{2}
\right) - f\left(\varepsilon-\frac{\hbar\omega}{2}\right) \right] .
\end{equation}
We have derived a new and, for our purposes, more convenient expression for the
diagonal part of the frequency-dependent transmission function,
\begin{eqnarray}
{\mathcal T}_{\mu\mu}(\varepsilon,\omega) &=& F_\mu\left( \frac{\hbar\omega}{2} -
\varepsilon - i 0^+, -\frac{\hbar\omega}{2} - \varepsilon - i 0^+ \right) \nonumber \\ &&
- F_\mu \left( \frac{\hbar\omega}{2} - \varepsilon - i 0^+, -\frac{\hbar\omega}{2} -
\varepsilon + i 0^+ \right) \nonumber \\ && \hspace{0.2cm} + \text{c.c.} \quad ,
\end{eqnarray}
given here in terms of functions
\begin{equation}
F_\mu (z, z^\prime) = \frac{\hbar \omega}{4} \sum_{\vek{r}} r_\mu^2 \,
\text{Tr}_{\text{sl}}\left\{ G_{\vek{r}}(z) G_{-\vek{r}} (z^\prime) \right\} \, .
\end{equation}
$G(\vek{r}, \vek{r^\prime}; z)\equiv G(\vek{r} - \vek{r^\prime},0; z) =: G_{\vek{r} -
\vek{r^\prime}}(z)$ is the real-space representation of the single-particle Greens
function in a translationally invariant system, and the trace $\text{Tr}_{\text{sl}}$ is
performed only over sublattice and layer degrees of freedom.

It is straightforward to specialize the general conductivity formula obtained in this
subsection to the clean limit.  Performing a Fourier transformation and using the fact
that the single-particle Hamiltonian $H\to H_{\vek{k}}$ becomes diagonal in real space,
we find
\begin{eqnarray}
F_\mu (z, z^\prime) &=& \frac{\hbar\omega}{16 \pi^2} \int \!\! d^2 k \,\,\,
\text{Tr}_{\text{sl}} \left\{ G_{\vek{k}}(z) \left[ \frac{\partial^2
H_{\vek{k}}}{\partial k_\mu^2} \right. \right. \nonumber \\ && \hspace{0.5cm} \left. \left.
- 2 \, \frac{\partial H_{\vek{k}}}{\partial k_\mu} \, G_{\vek{k}}(z) \, \frac{\partial
H_{\vek{k}}}{\partial k_\mu} \right] G_{\vek{k}}(z) G_{\vek{k}}(z^\prime)\right\} .
\end{eqnarray}
Here $G_{\vek{k}}(z) = (H_{\vek{k}} - z)^{-1}$ is the single-particle Greens function
in reciprocal-space (plane-wave) representation. Application to graphene yields the
same results as obtained more easily using formulae from the previous subsection.
However,  Eq.~(\ref{eq:newConduct}) turns out to be very useful beyond the clean
limit.

\section{Single-layer graphene in the clean limit}
\label{sec:singlay}

We apply the continuum-model description of a single sheet of graphene to evaluate
the conductivity formula (\ref{interTerm}). The single-particle Hamiltonian in plane-wave
representation is given by~\cite{sloncz:pr:58,Saito}
\begin{equation}
H_{\text{sg}} = \hbar v \left( k_x \sigma_x + k_y \sigma_y \right) + \tau \left[ \left( k_y^2
- k_x^2 \right) \sigma_x + 2 k_x k_y \sigma_y \right] ,
\end{equation}
where $v$ is the Dirac-fermion velocity characterizing the dispersion near the $\vek{K}$
point, and the term proportional to $\tau$ is a trigonal-warping correction to the band
structure. Straightforward diagonalization of $H_{\text{sg}}$ yields eigenvalues
$\varepsilon_{\vek{k}\sigma}^{\text{(sg)}}$, where $\sigma=\pm$ distinguishes the
electron and hole bands. As $\varepsilon_{\vek{k}\sigma}^{\text{(sg)}} \equiv -
\varepsilon_{\vek{k},-\sigma}^{\text{(sg)}}$, the dependence on temperature and
chemical potential is universal, i.e., independent of the values of $v$ and $\tau$. The
remaining dimensionless prefactor is only a function of $\omega$, $v$, $\tau$ and, by
simple dimensional analysis, can therefore only depend on these through the
combination $\omega\tau/(\hbar v^2)$. The expression for the conductivity reads then
\begin{equation}
\frac{\sigma_{\mu\nu}^{\text{(inter)}}}{\sigma_0} = g\left( \frac{\hbar\omega}{2 k_{\text{B}}
T}, \frac{\mu}{k_{\text{B}} T} \right) \,\, \Upsilon_{\mu\nu} \left(\frac{\omega\tau}{\hbar
v^2}\right) \,\, \Theta(\omega) \quad ,
\end{equation}
where $\Theta$ is the Heaviside step function, and we introduced the abbreviation
\begin{equation}
g(\xi, \eta) = \frac{\sinh\xi}{\cosh\xi + \cosh\eta} \quad .
\end{equation}
It is found that $\Upsilon_{xy}=\Upsilon_{yx}=0$ and
\begin{eqnarray}
&& \left. \begin{array}{c} \Upsilon_{xx}(\xi) \\ \Upsilon_{yy}(\xi) \end{array} \right\} =
2 \int \!\!  d^2\kappa \,\, \left\{ \begin{array}{l} \kappa_y^2 \left( 1 - 2 \xi \left[ \kappa_x
+ \xi \kappa^2 \right] \right)^2 \\ \left(1 + 2 \xi \, \kappa_x \right)^2 \left( \kappa_x - \xi
\kappa^2 \right)^2 \end{array} \right\} \nonumber \\
&& \hspace{1cm} \times \delta \left( 1 - 2 \sqrt{\kappa^2 - 2 \xi \, \kappa_x \left [ \kappa_x^2 - 3 \kappa_y^2 \right] + \xi^2 \kappa^4} \right) .
\end{eqnarray}
\begin{figure}[t]
\centerline{\includegraphics[width=3in]{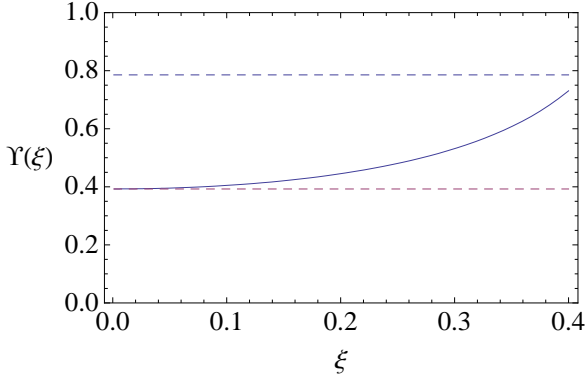}}
\caption{\label{fig:singTrigWarp}
Universal function capturing the effect of trigonal warping on the optical conductivity
of a single graphene sheet. 
The dashed lines indicate the values of $\pi/8$ and $\pi/4$, respectively.}
\end{figure}
We find numerically that $\Upsilon_{xx}(\xi)=\Upsilon_{yy}(\xi)\equiv \Upsilon(\xi)$ and
show this universal function in Fig.~\ref{fig:singTrigWarp}. In the limit $\tau\to 0$, the
well-known~\cite{ziegler:prb:07,falk:epjb:07,falk:08} universal conductivity of
single-layer graphene is found: $\Upsilon(0) = \pi/8$.
In the (for graphene unphysical) limit of large $\xi$, a different saturation value is realised:
$\Upsilon(\infty) = \pi/4$. 

\section{Bilayer graphene in the clean limit}
\label{sec:bilay}

\begin{figure}[t]
\centerline{\includegraphics[width=3in]{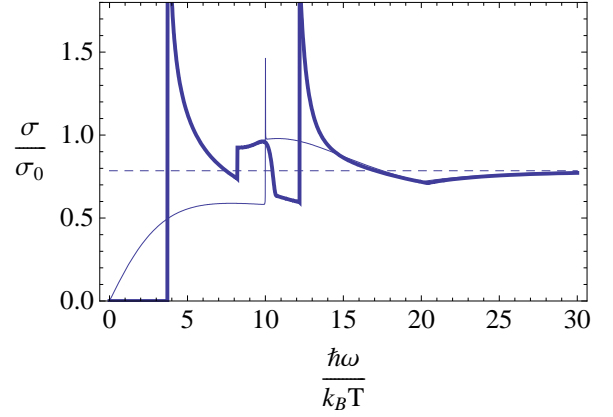}}
\caption{\label{fig:bilayer}
Interband contribution to the conductivity of clean bilayer graphene ($t_\perp = 10
k_{\text{B}}T$), with chemical potential at the symmetry point and a finite bias voltage
$V=2 k_{\text{B}}T/e$ between the layers (thick solid curve). The zero-bias case is
shown  as the thin solid curve. The dashed line indicates the value $\pi/4$.}
\end{figure}

To describe bilayer graphene, we use the $4 \times 4$ continuum-model
Hamiltonian~\cite{nilsson:prb:08} in plane-wave representation given by
\begin{equation}
H_{\text{bl}} = 
\begin{pmatrix}
e V/2 & \hbar v ( k_x + i k_y ) & t_{\perp} & 0\\
\hbar v ( k_x - i k_y ) & e V/2 & 0 & \hbar v_3 ( k_x + i k_y )\\
t_{\perp} & 0 & -e V/2 & \hbar v ( k_x - i k_y ) \\
0 & \hbar v_3 ( k_x - i k_y ) & \hbar v ( k_x + i k_y ) & -e V/2
\end{pmatrix}.
\end{equation}
Here $t_{\perp}$ and $V$ parameterize the strongest inter-layer hopping and a potential
difference applied between the layers, respectively. $v_3$ measures the strength of an
additional inter-layer hopping that gives rise to trigonal warping. Straightforward
diagonalization of $H_{\text{bl}}$ yields the set of energy eigenvalues
$\varepsilon_{\vek{k}\sigma}^{\text{(bl)}}$, with $\sigma=1, 2, 3, 4$. These eigenvalues
can be grouped into two pairs that add up to zero. Assuming without loss of generality
that $\varepsilon_{\vek{k}1}^{\text{(bl)}} < \varepsilon_{\vek{k}2}^{\text{(bl)}} \le \varepsilon_{\vek{k}3}^{\text{(bl)}} < \varepsilon_{\vek{k}4}^{\text{(bl)}}$, we have
$\varepsilon_{\vek{k}1}^{\text{(bl)}} = -\varepsilon_{\vek{k}4}^{\text{(bl)}}$ and
$\varepsilon_{\vek{k}2}^{\text{(bl)}} = -\varepsilon_{\vek{k}3}^{\text{(bl)}}$. This means
that there are {\em two\/} contributions of the type encountered in the single-layer case.
Depending on the situation, the remaining four contributions can be simplified as well.
Here we just give the analytical result obtained for the case with zero inter-layer
bias and trigonal warping neglected ($V=0$ and $v_3=0$). We find
$\sigma_{xx}^{\text{(inter)}}=\sigma_{yy}^{\text{(inter)}}\equiv \sigma_1+\sigma_2+
\sigma_3$, where
\begin{eqnarray}
\frac{\sigma_1}{\sigma_0} &=& \frac{\pi}{8} \,\, g\left( \frac{\hbar\omega}{2 k_{\text{B}} 
T}, \frac{\mu}{k_{\text{B}} T} \right) \left[ \frac{\hbar\omega + 2 t_\perp}{\hbar\omega +
t_\perp} \, \Theta(\omega) \right. \nonumber \\
&& \hspace{2cm} \left. + \frac{\hbar\omega - 2 t_\perp}{\hbar\omega - t_\perp} \, \Theta(\omega - 2 t_\perp/\hbar) \right] \, ,\\
\frac{\sigma_2}{\sigma_0} &=& \frac{\pi}{8} \left( \frac{t_\perp}{\hbar\omega} \right)^2
\Theta(\omega - t_\perp/\hbar) \left[ g\left( \frac{\hbar\omega}{2 k_{\text{B}} T}, 
\frac{2\mu - t_\perp}{2k_{\text{B}}T} \right) \right. \nonumber \\
&& \hspace{2.5cm} \left. + g\left(\frac{\hbar\omega}{2 k_{\text{B}} T}, \frac{2\mu +
t_\perp}{2k_{\text{B}} T}\right) \right] \, , \\
\frac{\sigma_3}{\sigma_0} &=& \frac{t_\perp}{\hbar}\, \delta(\omega - t_\perp/\hbar) 
\int_{\frac{t_\perp}{2 k_{\text{B}} T}}^\infty \frac{d\kappa}{\kappa} \left[ g\left(
\frac{t_\perp}{2 k_{\text{B}} T}, \frac{\mu}{k_{\text{B}}T} -\kappa \right) \right. \nonumber
\\ && \hspace{2.5cm} \left. + g\left(\frac{t_\perp}{2 k_{\text{B}} T}, \frac{\mu}{k_{\text{B}}
T} + \kappa\right) \right]  \, .
\end{eqnarray}
This result generalizes a previous expression~\cite{nicol:prb:08} obtained for the
zero-temperature limit. In Fig.~\ref{fig:bilayer}, we show the effect of a finite inter-layer
bias $V$.

\section{Conclusions}

We have studied the ac conductivity of single and bilayer graphene. Analytical results
were obtained for finite temperature and with trigonal warping included. Our
expressions should be useful to facilitate detailed comparison with experiment and
enable extraction of electronic-structure parameters from conductivity measurements.

\section*{Acknowledgment}

JZB is supported by a postdoctoral fellowship grant from the Massey University
Research Fund.



%% \bibliographystyle{elsarticle-num}
%% \bibliography{general,mesophys,graphene}

\end{document}